\documentclass[useAMS,usenatbib,usegraphicx]{mn2e} 
\usepackage{times}

\bibpunct{(}{)}{;}{a}{}{,}

\begin{document}

\title[]{TIRAVEL - Template Independent RAdial VELocity measurement}

\author[S. Zucker \& T. Mazeh]{
S.~Zucker$^1$ and T.~Mazeh$^2$ \\
$^1$Department of Geophysics and Planetary Sciences and Wise
  Observatory, Raymond and Beverly Sackler Faculty of Exact Sciences, \\
  Tel Aviv University, Tel Aviv, Israel \\
$^2$School of Physics and Astronomy, Raymond and Beverly
  Sackler Faculty of Exact Sciences, \\ 
  Tel Aviv University, Tel Aviv, Israel}

\maketitle
\begin{abstract}
We propose a new approach to measure differential radial velocities,
mainly for single-lined spectroscopic binaries. The proposed procedure
-- {\sc TIRAVEL} (Template Independent RAdial VELocities) -- does not
rely on a prior theoretical or observed template, but instead looks
for a set of relative Doppler shifts that simultaneously optimizes the
alignment of all the observed spectra. We suggest a simple measure to
quantify this overall alignment and use its maximum to measure the
relative radial velocities.  As a demonstration, we apply {\sc
TIRAVEL} to the observed spectra of three known spectroscopic
binaries, and show that in two cases {\sc TIRAVEL} performs as good as
the commonly used approach, while in one case {\sc TIRAVEL} yielded a
better orbital solution.
%
%
%
\end{abstract}

\section{Introduction}
\label{intro}

Since the seminal works of \citet{Sim1974} and \citet{TonDav1979}, the
cross-correlation technique to measure astronomical Doppler shifts has
become extremely popular.  The advent of digitized spectra and
computers rendered it the preferred method in astronomical fields
that require measurement of radial velocities (RVs) from observed
spectra, ranging from binary and multiple stellar systems to
cosmology. In recent years, improvements in the precision of RVs
measured through cross-correlation opened the way to the detection of many
extrasolar planets \citep[e.g.,][]{MayQue1995}.

The cross-correlation technique is conceptually simple and can be
presented in an intuitive manner. It searches for the velocity shift
of the observed stellar spectrum which maximizes the correlation with
a predetermined template. The template is assumed to adequately
represent the object spectrum with no Doppler shift.  The properties
of the cross-correlation technique have been studied extensively, in
an effort to improve its precision and overcome its few limitations,
and various methods have been suggested to estimate its precision
\citep[e.g.,][]{TonDav1979,Con1985,MurHea1991,Bouetal2001}.

The precision of the individual RVs depends on several factors, mainly
on the amount of spectral information and the signal-to-noise ratio
($S/N$) of the spectra. Another crucial factor is the quality of the
template spectrum, and the extent to which it represents the actual
stellar spectrum. Common practice is to use a theoretically calculated
spectrum as template \citep[e.g.,][]{Latetal2002}, a high $S/N$
exposure of another similar star or the observed star itself
\citep[e.g.,][]{Howetal1997}.

The theoretical synthetic spectrum might have some systematic
differences relative to the actual stellar
spectrum.  This happens not only because the theory of stellar
atmospheres is still not perfect and the spectral line lists are not
complete, but also because every star is somewhat different from other
stars, by its distinct abundances of the various atoms and ions in
particular. Similar problems occur when an observed spectrum is used
as template. An observed spectrum, either of the same star or of a
different star, introduces random noise to the template, thus
alienating it again from the true stellar spectrum.

We present here a new approach, {\sc TIRAVEL}, to derive the relative
velocity of stellar spectra without using any template. The approach
is based on the simple idea that when two spectra of the same star are
available, one can use cross-correlation to measure a {\it relative}
Doppler shift between the two spectra, each spectrum effectively
acting as a template for the other one.  {\sc TIRAVEL} is a
generalization of this approach for more than two spectra. We will
show that for large number of spectra, this approach is equivalent to
having a high-$S/N$ effective template.  Section \ref{tiravel}
introduces the approach, while Section \ref{realdata} applies it to
three real test cases. We discuss the test results and the
applicability of {\sc TIRAVEL} in Section \ref{discussion}.

\section{TIRAVEL}
\label{tiravel}

{\sc TIRAVEL} is basically a multi-spectral generalization of the
conventional cross-correlation technique. When we cross correlate two
spectra we scan a set of potential shifts between the spectra and
score them by the degree of mutual similarity between the two shifted
spectra. This score is based on the correlation coefficient, which has
been shown to be an optimal similarity measure under various
assumptions \citep[e.g.,][]{Zuc2003}.

Suppose we wish to estimate the relative Doppler shift between two
spectra $f_1(n)$ and $f_2(n)$.  The two spectra are described as
functions of the pixel number -- $n$, where $n = A\ln\lambda + B.$
Thus, the Doppler shift results in a uniform linear shift of the
spectrum \citep{TonDav1979}. For simplicity we assume the spectra were
already 'continuum-subtracted' and normalized, i.e.:
\begin{eqnarray}
\sum_n f_i(n) &= &0 \ , \\
\frac{1}{N} \sum_n f_i^2(n) &= &1 \ .
\end{eqnarray}
The cross-correlation function is then defined as: 
\[
R_{12}(s) = \frac{1}{N} \sum_n f_1(n) f_2(n-s) \ .  
\]
This expression is essentially the correlation coefficient between the
two spectra after they have been shifted according to the trial
relative shift $s$.  A higher value of $R_{12}(s)$ corresponds to
better agreement between the two sequences. Furthermore, it is well
known that $|R_{12}(s)| \le 1$.

{\sc TIRAVEL} generalizes the cross-correlation procedure to the case
of a set of $K>2$ spectra by considering the correlation matrix
$R_{ij}(s_1,s_2,\cdots,s_{\scriptscriptstyle K})$.  The entry $R_{ij}$
is the cross-correlation between the $i$-th and $j$-th spectrum, as a
function of their relative shift $s_i-s_j$. In order to measure the
overall agreement for a given set of shifts we propose to use
$\lambda_M$ -- the largest eigenvalue of the matrix
$R_{ij}(s_1,s_2,\cdots,s_{\scriptscriptstyle K})$.  We will show that
this estimate reduces to the correlation coefficient for $K=2$

The rationale behind this measure is simple and intuitive. At the
correct alignment, all the spectra are supposed to be easily modeled
by one {\it principal component}, or template. The highest eigenvalue
of the correlation matrix measures the degree to which we can describe
the spectra by one single principal component, as is well known from
Principal Component Analysis (PCA) theory
\citep[e.g.,][]{MurHec1987}. The eigenvector of $R_{ij}$, corresponding
to the eigenvalue $\lambda_M$, can be used to produce the principal
component itself, or in other words, the effective template {\sc
TIRAVEL} 'used' in order to produce the RVs. The elements of this
eigenvector, with proper normalization, are the optimal weights with
which the indvidual spectra should be summed to produce this
'template'. This is especially important in cases where the $S/N$
ratios are varying considerably among the observed spectra.

The value of the pairwise correlation coefficient can be interpreted
easily and intuitively, since values closer to unity mean better
agreement between the two sequences, and values close to zero mean
poorer agreement. We suggest to transform $\lambda_M$ to obtain
a similar behaviour:
\[
\rho =\frac{\lambda_M-1}{K-1} \ .
\]

{\sc TIRAVEL} essentially seeks the maximum of $\rho$ as a function of
the $K$ relative shifts $(s_1,s_2,\cdots,s_{\scriptscriptstyle K})$.
Each trial set of the $K$ relative shifts requires the calculation of
the corresponding correlation matrix $R_{ij}$. Therefore, an important
preliminary step in the computation should be the calculation of the
$K(K-1)/2$ cross-correlation functions, corresponding to all the pairs
of different spectra. The matrix elements $R_{ij}(s_i-s_j)$ are then
simply sampled from the pre-calculated cross-correlation functions.

Note also that the elements of the correlation matrix $R_{ij}$ depend
only on pairwise differences of shifts, i.e., relative shifts. It is
therefore insensitive to zero-point shifts which are common to all the
observed spectra. In that respect, the shifts derived by {\sc TIRAVEL}
are only relative shifts, with effective $K-1$ degrees of freedom
instead of $K$.

After optimizing over $(s_1,s_2,\cdots,s_{\scriptscriptstyle K})$,
which can be done by any optimization procedure
\citep[e.g.,][]{Preetal1992}, we obtained the final velocities by
fitting parabolas to the peak, similarly to the common practice in
regular cross-correlation.

Since the new procedure yields also an effective 'template', produced
by summing all the individual exposures at the right shifts with
optimal weights, we expect the quality of this effective template to
improve as the number of observed spectra increases. Better template
should yield more accurate individual velocities. We therefore expect
the precision of the {\sc TIRAVEL} velocities to improve as a function
of the number of spectra. To test this we ran {\sc TIRAVEL} on
different sets of simulated spectra and checked the resulting
velocities. We generated $K$ spectra by using a synthetic spectrum of
a G star to which we added normal noise at a specified $S/N$. No shift
was implemented in producing the spectra, and therefore the scatter of
the velocities derived by {\sc TIRAVEL} can serve as a measure of the
precision of the approach. We repeated the procedure $50$ times for
each $S/N$ and $K$.

\begin{figure}
\resizebox{\hsize}{!}{\includegraphics{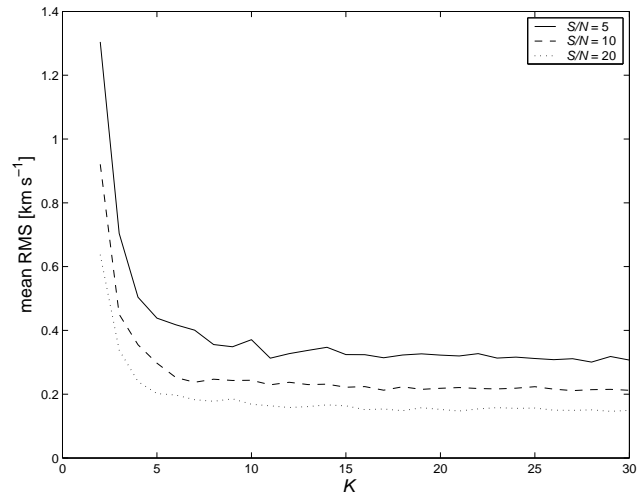}}
\caption{Results of applying {\sc TIRAVEL} to sets of $K$ simulated 
spectra with zero Doppler shift. The plots present the mean RMS of the
velocities produced by {\sc TIRAVEL}.}
\label{Kplot}
\end{figure}

Figure \ref{Kplot} displays the results of these simulations.  The
plot shows the mean RMS of the velocities obtained by {\sc TIRAVEL} as
a function of the number of spectra analysed, for three different
$S/N$-ratios. As expected, the RMS converges to a constant value, when
the 'effective template' becomes similar to the true template. It
seems that the convergence rate is not very sensitive to the $S/N$,
and that when $10$--$15$ spectra are available, the ultimate
performance is already achieved (assuming they all share about the
same $S/N$). This behaviour is easily understood, since once the
template is significantly less noisier than the individual spectra,
the main factor which determines the velocity precision is the $S/N$
of the observed spectrum. This happens already when a moderate number
of spectra are co-added to produce the template. 

The simulations also indicate the computational burden of {\sc
TIRAVEL}. Using a very na\"{\i}ve optimization scheme, on average
about $10$ iterations were needed to reach the stopping criterion of
less than $0.01\,\mathrm{km\ s}^{-1}$ RMS change in the velocities,
for $K=30$. We wrote our code in {\sc MATLAB} and ran it on a Pentium
Pro Linux machine, where those $10$ iterations, including the
pre-calculation step, lasted about $20$ seconds.

\section{Real Test Cases}
\label{realdata}

We chose to demonstrate the potential of the new approach on three
single-lined spectroscopic binaries (SB1s) which were already solved
and published \citep{Latetal2002} as part of a survey of $171$
high-proper-motion SB1s.  The radial velocities in \citet{Latetal2002}
were derived from spectra obtained using the CfA Digital Speedometers
\citep{Lat1985,Lat1992}. In order to obtain the radial velocities
\citeauthor{Latetal2002} used an extensive grid of spectra computed
using the model atmospheres of the {\sc ATLAS9} code, developed by
Kurucz \citetext{Morse \& Kurucz, in preparation}.

The spectra we used for the {\sc TIRAVEL} analysis were the same
observed spectra that \citeauthor{Latetal2002} used for their
analysis.  Using the velocities produced by {\sc TIRAVEL} we have
recalculated a least-square best-fitting orbital solution.  The RMS of
the residuals of the new orbit serves as a criterion to compare with
the previously published orbit. Furthermore, the effective template
derived by {\sc TIRAVEL} indicates the amount of spectral information
that can be extracted from the spectra without using a template.

\subsection{G72--59}
\label{G72-59}

\begin{figure}
\resizebox{\hsize}{!}{\includegraphics{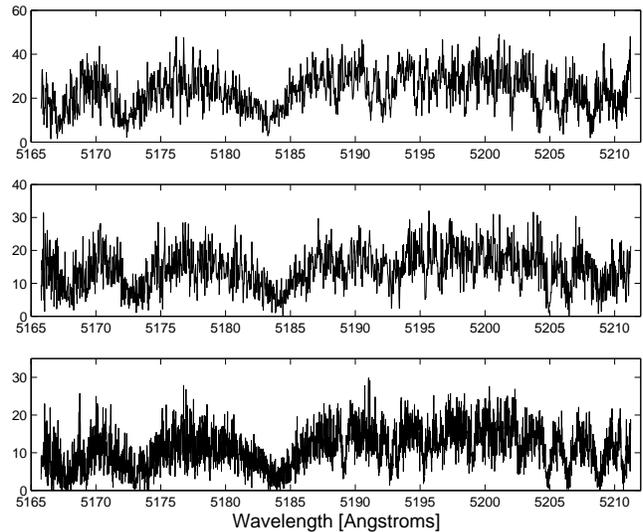}}
\caption{Sample spectra of G72--59. The  times of the presented exposures are 
$\rmn{JD}=2447053.7466$ (upper panel), $\rmn{JD}=2447163.5952$ (middle panel) 
and $\rmn{JD}=2449678.7457$ (lower panel).}
\label{G72-59_specs}
\end{figure}

\citet{Latetal2002} solved G72--59 with 32 spectra
and derived a period of about $88$ days with a small
eccentricity. Figure \ref{G72-59_specs} shows three of the 32 spectra,
to demonstrate their S/N. Table \ref{G72-59_table} lists both the
orbital elements that appear in \citet{Latetal2002} and the orbital
elements obtained from the {\sc TIRAVEL} velocities.  Note that the {\sc
TIRAVEL} analysis reveals no information about {\it absolute} radial
velocities, and the value of the center-of-mass velocity obtained in
analysing {\sc TIRAVEL} velocities is not very
informative. Nevertheless we list it for completeness.

\begin{figure}
\resizebox{\hsize}{!}{\includegraphics{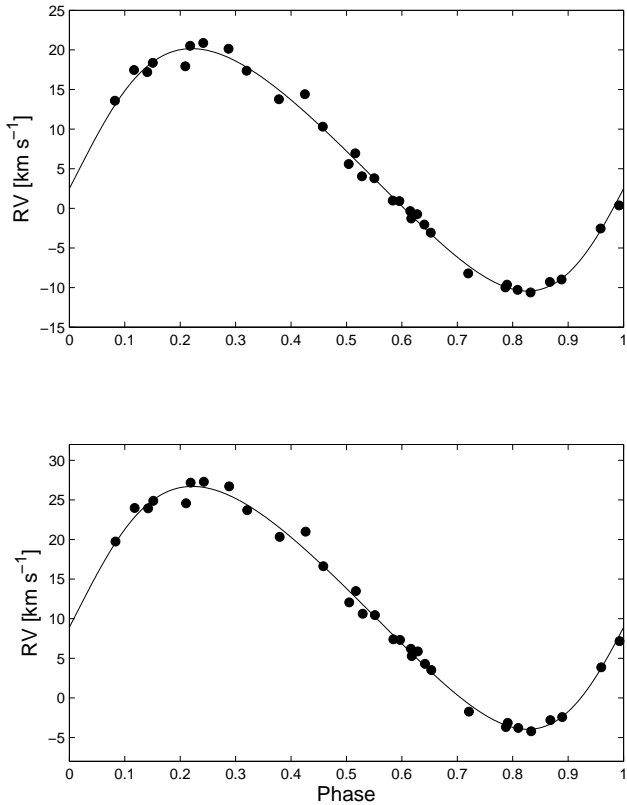}}
\caption{Upper panel: The orbital solution of G72--59 using 
{\sc TIRAVEL} velocities. 
Lower panel: The old orbital solution by \citet{Latetal2002}.}
\label{G72-59_orbits}
\end{figure}

\begin{figure}
\resizebox{\hsize}{!}{\includegraphics{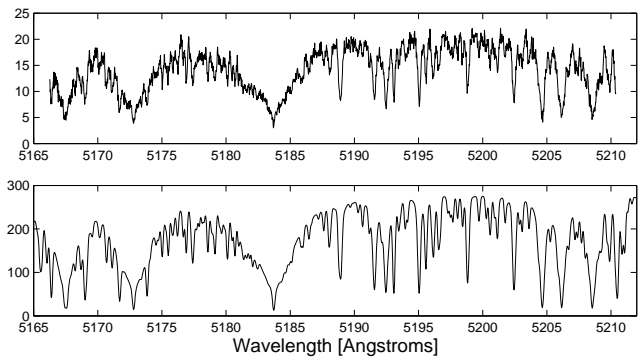}}
\caption{Upper panel: The effective template spectrum obtained by 
{\sc TIRAVEL} for 
the spectra of G72--59. Lower panel: the Kurucz spectrum used as
template by \citet{Latetal2002}, shifted to match the {\sc TIRAVEL}
effective template.}
\label{G72-59_temps}
\end{figure}

Table \ref{G72-59_table} shows that the two solutions are
consistent. The residual RMS of the {\sc TIRAVEL} solution is somewhat
larger than the old one, and so are the errors of the elements. This
is one of the cases where the {\sc TIRAVEL} solution is a little worse
than \citet{Latetal2002} solution.  The two orbital solutions can be
visually compared in Figure \ref{G72-59_orbits}. Indeed, the
differences between the two orbits are completely negligible. Figure
\ref{G72-59_temps} presents the new effective template that is built
by {\sc TIRAVEL}, together with the synthetic template used in
the original analysis.

The synthetic template shown in Figure \ref{G72-59_temps} was shifted
according to the difference between the two derived center-of-mass
velocities, in order to match the derived {\sc TIRAVEL} template. The
similarity is striking, although the synthetic template evidently has
a much higher $S/N$ than the derived one. It is somewhat surprising
and gratifying that the {\sc TIRAVEL} results are almost as good as
the results of the calculated spectrum, despite the fact that the
effective spectrum of {\sc TIRAVEL} looks so much noisier.

\begin{table}
\caption{Comparing the two orbital solutions for G72--59}
\begin{tabular}{lc*{2}{r@{\,$\pm$\,}l}}
Parameter & &
\multicolumn{2}{c}{\citealt{Latetal2002}} & 
\multicolumn{2}{c}{\sc TIRAVEL} \\
\hline
$P$      & [days]         & $87.754$        & $0.026$ & $87.755$        & $0.031$ \\ 
$T$      & [JD]           & $2\,447\,701.7$ & $1.4$   & $2\,447\,701.8$ & $1.5$   \\ 
$e$      &                & $0.168$         & $0.017$ & $0.169$         & $0.019$ \\
$\omega$ & [$\degr$]      & $260.8$         & $5.4$   & $261.1$         & $6.1$   \\
$K$      & [km\ s$^{-1}$] & $15.33$         & $0.24$  & $15.30$         & $0.28$  \\
$V_0$    & [km\ s$^{-1}$] & $11.78$         & $0.18$  & $5.26$          & $0.20$  \\
\hline
$\sigma_{\scriptscriptstyle O-C}$ & [km\ s$^{-1}$] & \multicolumn{2}{c}{$0.94$} & \multicolumn{2}{c}{0.96}
\end{tabular}
\label{G72-59_table}
\end{table}
          
\subsection{G178--27}
\label{G178-27}

\begin{figure}
\resizebox{\hsize}{!}{\includegraphics{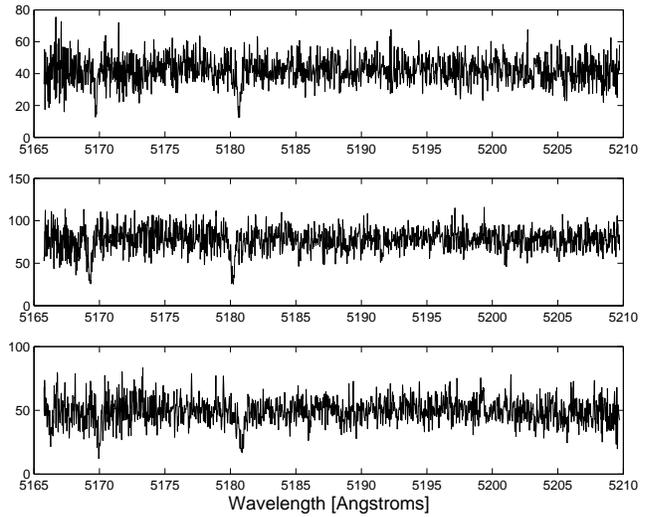}}
\caption{Sample spectra of G178--27. The  times of the presented exposures are 
$\rmn{JD}=2446928.9137$ (upper panel), $\rmn{JD}=2447494.0177$ (middle
panel) and $\rmn{JD}=2447702.7553$ (lower panel).}
\label{G178-27_specs}
\end{figure}

According to \citet{Latetal2002}, G178--27 has a low amplitude orbit
($K = 3.4$\,km\ s$^{-1}$), with an eccentricity of about $0.4$. Its
period is about $81$ days, and the orbital solution was obtained using
$31$ exposures. \citeauthor{Latetal2002} derived very low
metallicity for G178--27, of $[\rmn{Fe}/\rmn{H}] = -2.0$. Figure
\ref{G178-27_specs} presents three sample spectra of G178--27.  One
can see the relative scarcity of the spectral information compared with
G72--59, due to the very low metallicity.

\begin{figure}
\resizebox{\hsize}{!}{\includegraphics{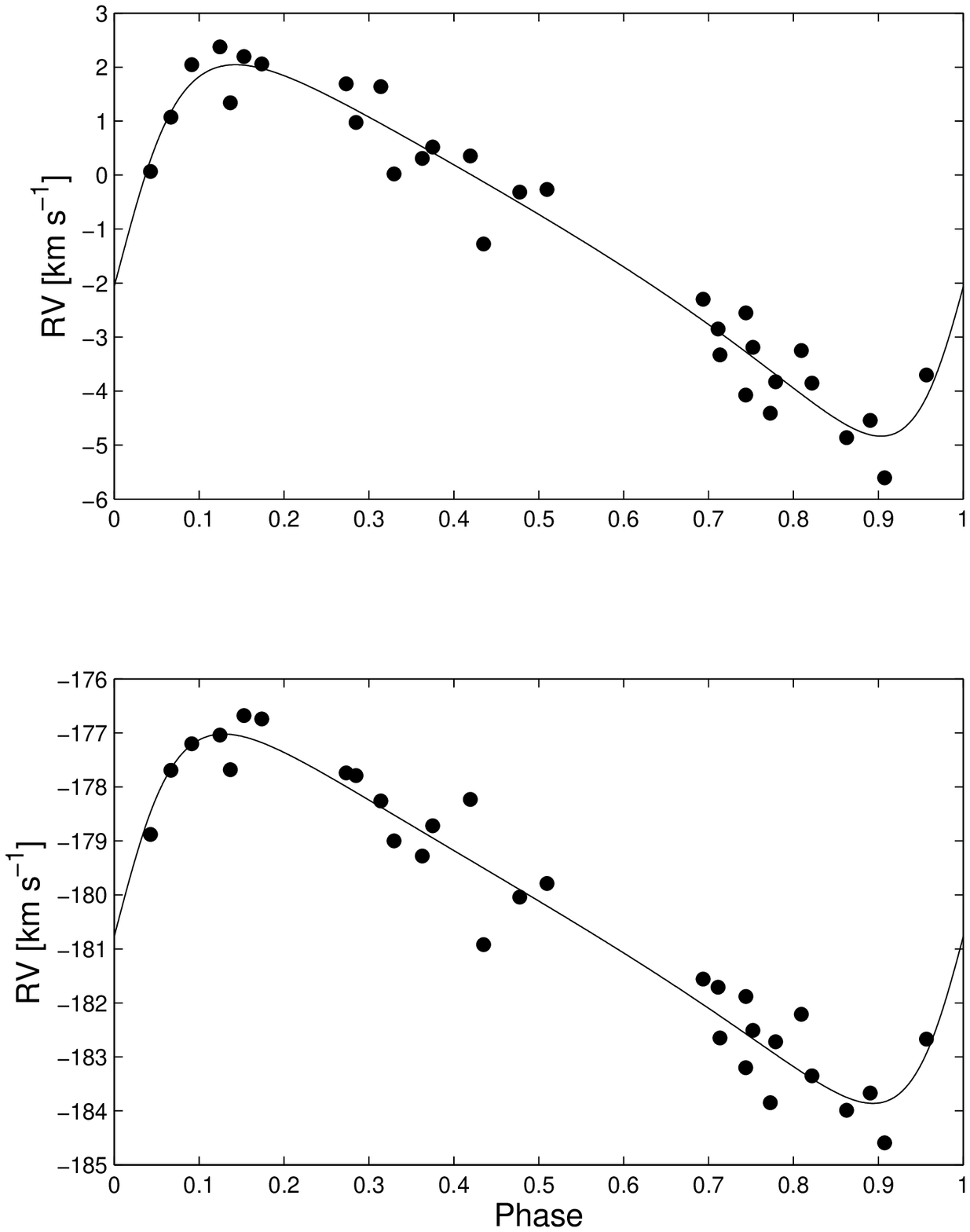}}
\caption{Upper panel: The orbital solution of G178--27 using 
{\sc TIRAVEL} velocities. 
Lower panel: The previous orbital solution by \citet{Latetal2002}.}
\label{G178-27_orbits}
\end{figure}

\begin{figure}
\resizebox{\hsize}{!}{\includegraphics{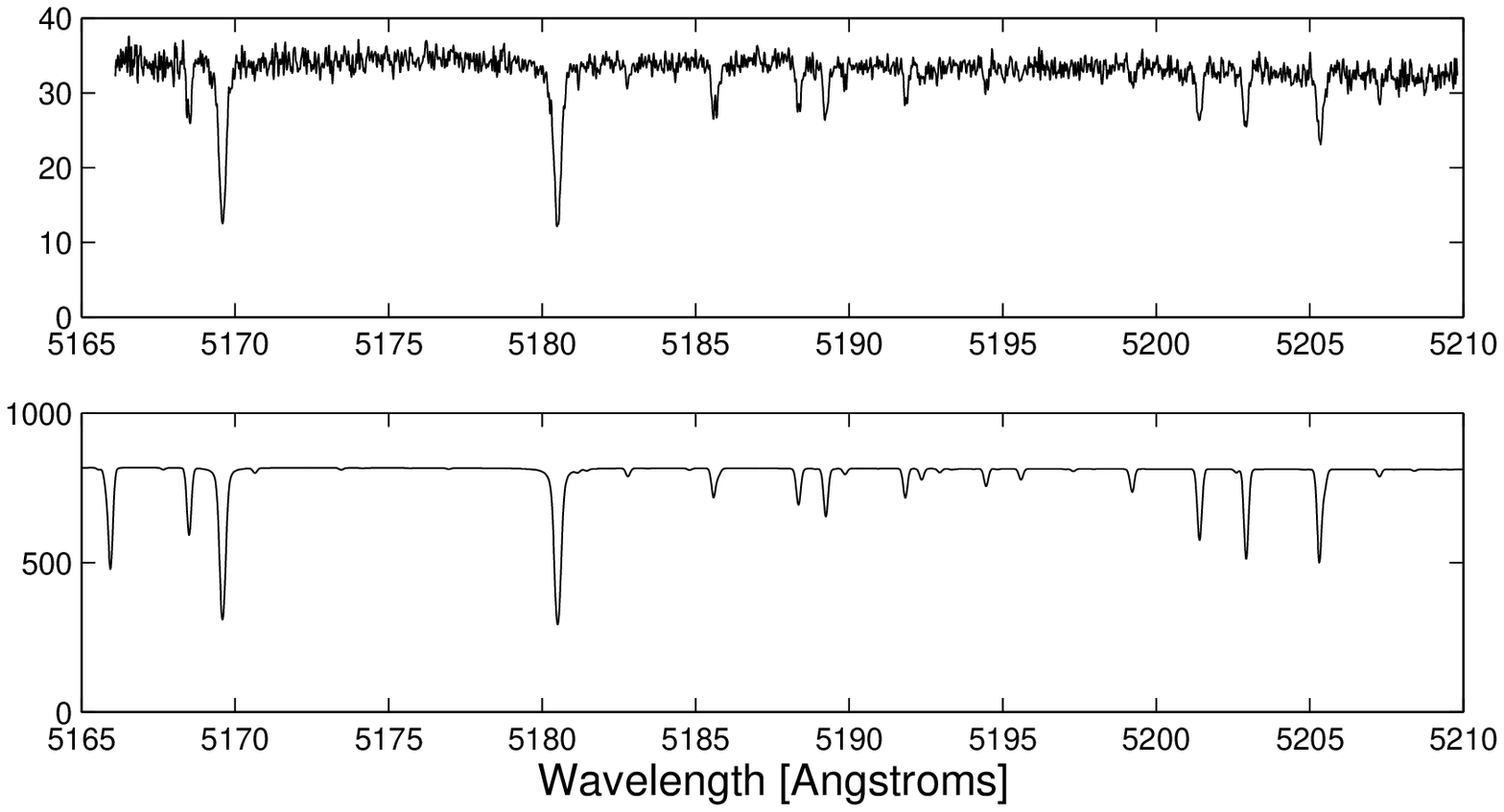}}
\caption{Upper panel: The effective template spectrum obtained 
by {\sc TIRAVEL} for 
the spectra of G178--27. Lower panel: the Kurucz spectrum used as
template by \citet{Latetal2002}, shifted to match the {\sc TIRAVEL}
effective template}
\label{G178-27_temps}
\end{figure}

The two orbital solutions are listed in Table \ref{G178-27_table}, and
shown on Figure \ref{G178-27_orbits}. Once again, the new solution is
very similar and consistent with the previous one, while its residuals are
slightly smaller.  The good choice of synthetic
template by \citeauthor{Latetal2002} is clearly seen once the {\sc
TIRAVEL} effective template is derived, as shown in Figure
\ref{G178-27_temps}.

\begin{table}
\caption{Comparing the two orbital solutions for G178--27}
\begin{tabular}{lc*{2}{r@{\,$\pm$\,}l}}
Parameter & &
\multicolumn{2}{c}{\citealt{Latetal2002}} & 
\multicolumn{2}{c}{\sc TIRAVEL} \\
\hline
$P$      & [days]         & $81.18$         & $0.37$  & $81.18$         & $0.35$  \\ 
$T$      & [JD]           & $2\,447\,436.5$ & $1.7$   & $2\,447\,436.1$ & $1.8$   \\ 
$e$      &                & $0.431$         & $0.046$ & $0.428$         & $0.045$ \\
$\omega$ & [$\degr$]      & $264.5$         & $8.5$   & $258.8$         & $9.1$   \\
$K$      & [km\ s$^{-1}$] & $3.42$          & $0.18$  & $3.44$          & $0.18$  \\
$V_0$    & [km\ s$^{-1}$] & $-180.30$       & $0.12$  & $-1.11$         & $0.13$  \\
\hline
$\sigma_{\scriptscriptstyle O-C}$ & [km\ s$^{-1}$] & \multicolumn{2}{c}{$0.61$} & \multicolumn{2}{c}{0.56}
\end{tabular}
\label{G178-27_table}
\end{table}

\subsection{G48--54}
\label{G48-54}

\begin{figure}
\resizebox{\hsize}{!}{\includegraphics{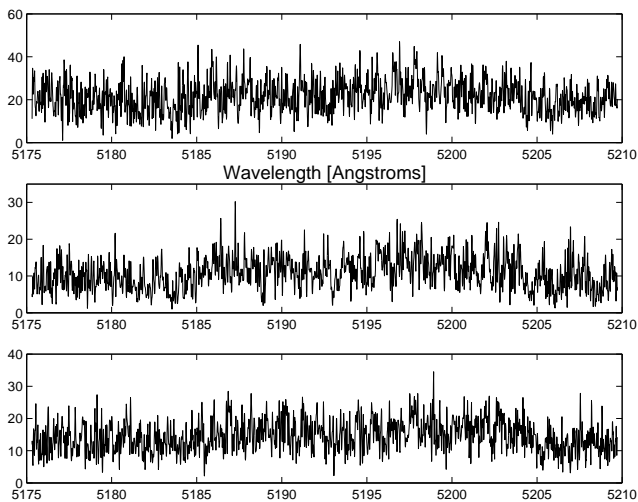}}
\caption{Sample spectra of G48--54. The  times of the presented exposures are 
$\rmn{JD}=2445770.8684$ (upper panel), $\rmn{JD}=2446861.8126$ (middle
panel) and $\rmn{JD}=2446959.6461$ (lower panel).}
\label{G48-54_specs}
\end{figure}

The star G48--54 is the best case considered here to demonstrate the
potential of {\sc TIRAVEL}. The previous solution by
\citeauthor{Latetal2002} has a period of $22.6$ days and an
eccentricity of $0.3$, obtained with $30$ exposures. The star G48--54
is the star with the latest spectral type presented in
\citet{Latetal2002}, with an estimated temperature of
$3750\,\rmn{K}$. Figure \ref{G48-54_specs} presents three sample
spectra of G48--54.  

\begin{figure}
\resizebox{\hsize}{!}{\includegraphics{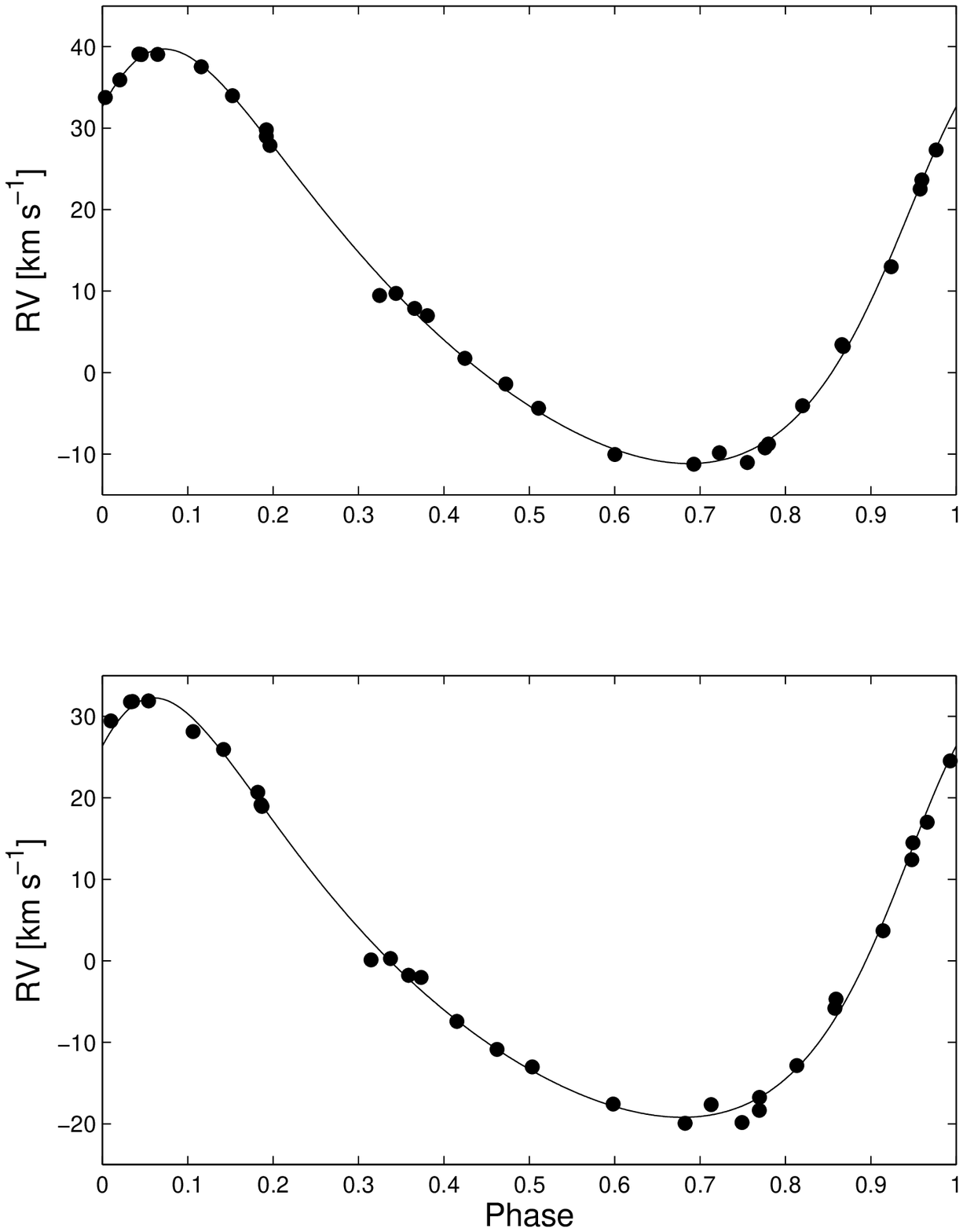}}
\caption{Upper panel: The orbital solution of G48--54 
using {\sc TIRAVEL} velocities. 
Lower panel: The previous orbital solution by \citet{Latetal2002}.}
\label{G48-54_orbits}
\end{figure}

\begin{figure}
\resizebox{\hsize}{!}{\includegraphics{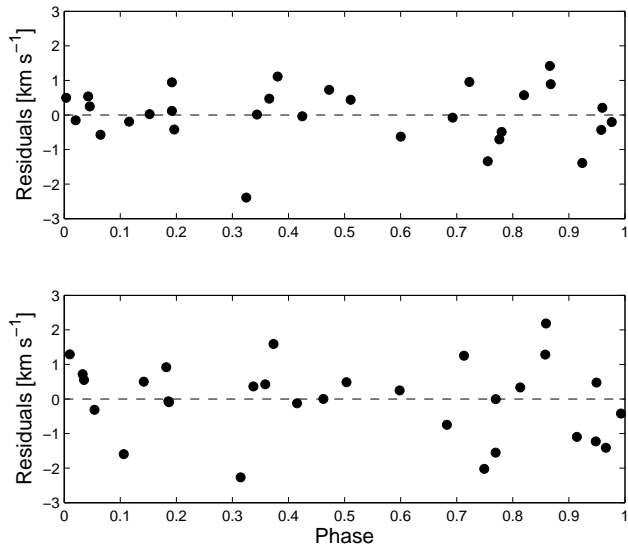}}
\caption{Upper panel: The residual RVs after subtracting the 
best-fitting orbital solution from the {\sc TIRAVEL} velocities of
G48--54. Lower panel: The residual RVs after subtracting the previous
best-fitting orbital solution from the velocities published by
\citet{Latetal2002}.}
\label{G48-54_residuals}
\end{figure}

\begin{figure}
\resizebox{\hsize}{!}{\includegraphics{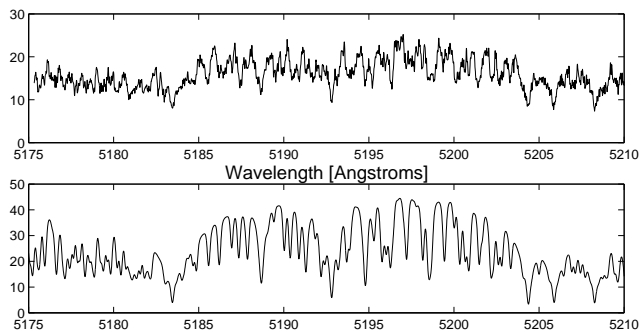}}
\caption{Upper panel: The effective template spectrum obtained 
by {\sc TIRAVEL} for 
the spectra of G48--54. Lower panel: the Kurucz spectrum used as
template by \citet{Latetal2002}, shifted to match the {\sc TIRAVEL}
effective template}
\label{G48-54_temps}
\end{figure}

As can be seen in Table \ref{G48-54_table}, the {\sc TIRAVEL} solution
is better than the old published solution. Because of the large RV
amplitude, the differences between the two solutions can hardly be
seen in the phase-folded RV curve (Figure \ref{G48-54_orbits}). We
therefore present in Figure \ref{G48-54_residuals} the residuals after
subtracting the orbital solution in the two cases, on the same scale.
The improvement is easily seen. Figure \ref{G48-54_temps} suggests the
reason for the improvement. The spectrum of this late-type star is
more complex than most of the other stars in the sample, because of
its lower temperature. We suggest that the spectral modeling of this
star is not
complete, probably due to the notorious difficulties associated with
modeling later-type stars.

Close examination of Figures \ref{G48-54_orbits} and
\ref{G48-54_residuals} suggests that a single measurement, at a phase
of about $0.3$, may be 'pulled' by the secondary spectrum, at a phase
where the primary and secondary velocities are very close. This
outlying velocity stands out even better in the {\sc TIRAVEL}
velocities, because of the smaller RMS. Maybe a somewhat better
procedure would exclude this point from the analysis.

\begin{table}
\caption{Comparing the two orbital solutions for G48--54}
\begin{tabular}{lc*{2}{r@{\,$\pm$\,}l}}
Parameter & &
\multicolumn{2}{c}{\citealt{Latetal2002}} & 
\multicolumn{2}{c}{\sc TIRAVEL} \\
\hline
$P$      & [days]         & $22.6057$        & $0.0049$ & $22.6020$        & $0.0042$ \\ 
$T$      & [JD]           & $2\,446\,729.39$ & $0.16$   & $2\,446\,729.19$ & $0.15$   \\ 
$e$      &                & $0.287$          & $0.012$  & $0.257$          & $0.010$  \\
$\omega$ & [$\degr$]      & $320.4$          & $2.8$    & $316.2$          & $2.6$    \\
$K$      & [km\ s$^{-1}$] & $25.74$          & $0.32$   & $25.44$          & $0.27$   \\
$V_0$    & [km\ s$^{-1}$] & $0.84$           & $0.24$   & $9.55$           & $0.20$   \\
\hline
$\sigma_{\scriptscriptstyle O-C}$ & [km\ s$^{-1}$] & \multicolumn{2}{c}{$1.20$} & \multicolumn{2}{c}{0.89}
\end{tabular}
\label{G48-54_table}
\end{table}

\section{Discussion}
\label{discussion}

We have presented here a new technique to analyse the spectra of SB1
systems. {\sc TIRAVEL} is independent of any assumption about the
primary spectrum, which is not included in the actual observed
data. Instead, {\sc TIRAVEL} makes use of the observed spectra to
derive the radial velocities of the primary. This approach assumes
that there is only one component in the observed spectra and that they
are free of atmospherical or mechanical systematic features. Such
features might reduce the efficiency of the technique, in particular
compared to the use of theoretically calculated templates.

We have applied {\sc TIRAVEL} to three SB1 cases, where a satisfactory
solution had already obtained with theoretically calculated
templates. {\sc TIRAVEL} improved significantly the quality of one of
the three solutions, in terms of the precision of the orbital elements
and the RMS residuals. Obviously, the advantages of {\sc TIRAVEL} will
be most pronounced when there is no good theoretically calculated
template available. In the presented cases, the template grid that
\citet{Latetal2002} used was very extensive, and we could not expect
to obtain very large improvements, except at the low-temperature
frontier of the grid. One immediate implication of {\sc TIRAVEL} is
the analysis of the spectra of cataclysmic variables or some T Tauri
stars, where the high-temperature accretion disc has an emission
spectrum which in not easy to model.

The version of {\sc TIRAVEL} we presented here is specifically
tailored for SB1s, where there is only one RV to be measured in each
exposure, resulting in a simple shift of the spectrum. In the case of
double-lined spectroscopic binaries (SB2), a more elaborate scheme is
needed, one that has to separate the effects of the orbit on the two
component spectra.  Such a scheme may make use of ideas related to
Doppler tomography \citep{Bagetal1992} or two-dimensional correlation
({\sc TODCOR};\citet{ZucMaz1994}). One possible generalization of {\sc
TIRAVEL} may simply apply it iteratively - first solving for the
primary component, subtracting the resulting template from the
spectra, then applying {\sc TIRAVEL} to the subtracted spectra again
to retrieve the information related to the secondary component, and
repeat the process iteratively. Similar procedures were applied by
\citet{Maretal1998} and \citet{GonLev2006}.

Note that spectral disentangling codes, like {\sc KOREL}
\citep{Had1995,Had1997} can be applied also to SB1s
\citep[e.g.,][]{Saaetal2005}. However, they
usually impose an orbital model, treating any deviation from the model
as noise.  {\sc TIRAVEL} does not incorporate any orbital model (or
any other source of RV variation), and produces the best relative RV
estimates one can get without a template. Therefore, the {\sc TIRAVEL}
approach enables us to detect any additional source for RV variation,
a third companion in particular.

In order to be applicable to spectra obtained by the modern echelle
spectrographs and other multi-order spectra, {\sc TIRAVEL} has to be
generalized, probably along lines similar to the ones that were used
to generalize {\sc TODCOR} to multi-order spectra
\citep{Zuc2003}. This modification was crucial in detecting the planet
in HD\,41004 \citep{Zucetal2003,Zucetal2004}.

In principle, improvements in RV precision, like the one offered by
{\sc TIRAVEL}, may lead to the detection of small finite
eccentricities or additional stellar components. One area where the
precision of radial velocities and orbital solution is of utmost
importance is the search for extrasolar planets, where the orbital RV
amplitude is usually extremely small, of the order of tens of meters
per second. Effective improvements in the precision of radial
velocities may lead to an improved planet-detection capability. In
general, the improved precision also means a better utilization of the
instrumental potential and thus a more efficient observing time usage.

\section*{Acknowledgments}

We are deeply indebted to Dave Latham for his most useful
assistance in accessing and analysing the CfA spectra. 

\bibliographystyle{mn2e}
\bibliography{ref}

\end{document}